# Chemical Synthesis and Magnetic Properties of Monodisperse Cobalt Ferrite Nanoparticles

Z. Mahhouti[1,2,3*], H. El Moussaoui[1], T. Mahfoud[1], M. Hamedoun[1], M. El Marssi[3], A. Lahmar[3], A. El Kenz[2], and A. Benyoussef[1,2,4]

[1] MAScIR Foundation, Institute of Nanomaterials and Nanotechnologies, Materials and Nanomaterials Center, B.P. 10100, Rabat, Morocco

[2] LaMCScI, URAC 12, Département de Physique, Faculté des Sciences, Université Mohammed V, B.P. 1014, Rabat, Morocco

[3] LPMC EA2081, Université de Picardie Jules Verne 33 Rue Saint Leu, 80000 Amiens, France

[4] Hassan II Academy of Science and Technology, Rabat, Morocco

**Abstract**

In this work, a successful synthesis of magnetic cobalt ferrite ($CoFe_2O_4$) nanoparticles is presented. The synthesized $CoFe_2O_4$ nanoparticles have a spherical shape and highly monodisperse in the selected solvent. The effect of different reaction conditions such as temperature, reaction time and varying capping agents on the phase and morphology is studied. Scanning transmission electron microscopy showed that the size of these nanoparticles can be controlled by varying reaction conditions. Both X-ray diffraction and energy dispersive X-ray spectroscopy corroborate the formation of $CoFe_2O_4$ spinel structure with cubic symmetry. Due to optimized reaction parameters, each nanoparticle was shown to be a single magnetic domain with diameter ranges from 6 nm to 16 nm. Finally, the magnetic investigations showed that the obtained nanoparticles are superparamagnetic with a small coercivity value of about 315 Oe and a saturation magnetization of 58 emu/g at room temperature. These results make the cobalt ferrite nanoparticles promising for advanced magnetic nanodevices and biomagnetic applications.

# 1. Introduction

The magnetic nanoparticles with spinel structure $MFe_2O_4$ (M = Fe, Co, Mn, Zn, Ni ...) have been widely studied for their properties compatible with various applications ranging from data storage to biomedical applications [1, 2, 3, 4, 5]. Recently, a special interest is devoted to magnetic nano-object materials [6, 7], because they endorsed interesting magnetic properties, with the possibility of tailoring their functionalities, by controlling the shape and morphology. Particularly, magnetic nanoparticles (MNPs) can be tuned in a straight forward manner by the control of the size, monodispersity, chemical composition, as well as the adequate synthesis route, which is desirable for advanced magnetic nanodevices or magnetic hyperthermia.

It is worthy to mention that a monodomain nanoparticle has a permanent magnetic moment, which is the sum of all magnetic moments of the atoms constituting it. However, during the structuring of the magnetic monodomain, the reduction of the total number of atoms (on the nanometric scale) leads to an increase in the contribution of surface atoms that do not have the same environment as in the core of the nanoparticle.

The critical diameter $d_C$ from which the particle can be considered as a magnetic monodomain is defined by Frey et al. [1]:

$$d_C = \frac{36\sqrt{AK_{eff}}}{M_s^2 \mu_0} \qquad (Eq.1)$$

Where $K_{eff}$ is the effective anisotropy and $A$ is the exchange constant. $\mu_0$ is the vacuum permeability and $M_S$ is the saturation magnetization. $d_C$ is in the range of 10-100 nm.

The contribution of surface effects affect the magnetic properties of the material [8]. Indeed, in addition to the core spins as in the bulk material, the nanoparticles have surface spins, that creating supplementary interactions. Therefore, for controlling the physical and chemical properties of nanoparticles, it is necessary to control the size, morphology, monodispersity, and chemical composition of the nanoparticles.

For instance, cobalt ferrite has a ferromagnetic behavior at ambient conditions, with high magnetic coercivity [8], a high Curie temperature at the vicinity of 793 K, strong magnetocrystalline anisotropy [8, 9], as well as a large magnetostriction coefficient [10]. These properties are very attracting for advanced technological devices, namely in data storage and in the biomedical applications. Magnetic order in cobalt ferrite arises from the superexchange interaction between the cations located in tetrahedral and octahedral sites through the oxygen anion. The induced antiferromagnetic coupling between the $Fe^{3+}$ cations in the tetrahedral sites and the $Co^{2+}$ and $Fe^{3+}$ cations in the octahedral sites is strong; although another weak antiferromagnetic coupling is present between tetrahedral $Fe^{3+}$ cations [11, 12]. In addition, a weak ferromagnetic coupling also exists between the cations of the octahedral sites. The two last couplings are masked by the interactions between tetrahedral and octahedral sites.

To date, several synthesis methods of MNPs have been developed [13, 14, 15, 16, 17, 18] in an effort to improve the magnetic properties by controlling the size, the morphology, and the composition of the obtained nanoparticles. Among these different routes of synthesis, we have found the co-precipitations, solvothermal, hydrothermal and thermal decomposition which are the most effective ones. The co-precipitation method has been used to synthesize crystals with different morphologies including spherical, cubic and nanorods [19]. Using solvothermal and hydrothermal methods, nanocrystals of iron oxide have been grown as spheres and hexagons [16, 17]. Thermal decomposition method has produced monodispersed nanoparticles of spinel ferrite with a narrow size distribution and good crystallinity [22]. In this respect, the present work reports on the synthesis of $CoFe_2O_4$ NPs by decomposition of acetylacetonate precursors at high temperature. Among many advantages of this synthesis route, the ability to control the particle size, size distribution, shape, and phase purity. The thermal decomposition approach has been chosen because the synthesis system is simple with one type of complexes, one type of ligands and a high boiling point organic solvent. The obtained nanoparticles are monodisperse with varied morphologies and sizes.

The $CoFe_2O_4$ nanoparticles synthesized with the standard protocol were characterized using many experimental techniques such as ThermoGravimetric Analysis (TGA), X-Ray Diffraction (XRD), Fourier Transform InfraRed spectroscopy (FT-IR), Scanning Transmission Electron Microscopy (STEM), Energy Dispersive X-ray Spectroscopy (EDS) and Magnetic Property Measurement System (MPMS) SQUID magnetometer.

Our main research topic in this work is especially the development of low cost, flexibility, and ease of chemical synthesis of $CoFe_2O_4$ nanoparticles (NPs). To deep understanding how some synthesis parameters affect the nucleation and growth steps; the decomposition temperature, reflux time, nature of solvents, the quantity of surfactants were investigated. Therefore, the $CoFe_2O_4$ nanoparticles obtained by varying the experimental conditions were characterized by STEM in order to describe the influence of synthesis parameters on the size and shape of NPs.

## 2. Experimental section

### 2.1. Chemicals

The synthesis was carried out using commercially available reagents. The starting Precursors were iron(III) acetylacetonate (Fe(acac)$_3$, 99.99%), and cobalt(II) acetylacetonate (Co(acac)$_2$, 99%). The used solvents were Benzyl ether (98%, boiling point: 298 °C), absolute ethanol (100%), and hexane (98.5%). For the surfactants and reductant we used oleic acid (90%, boiling point: 360 °C), oleylamine (70%, boiling point: 350 °C), and 1,2-hexadecanediol (90%). All the chemicals were purchased from Sigma-Aldrich Ltd. and were used as received without further purification.

### 2.2. Synthesis of cobalt ferrite nanoparticles

Into a 100 mL three-necked flask under nitrogen flow, we placed 4 mmol of Fe(III) acetylacetonate, 2 mmol of Co(II) acetylacetonate, 20 mmol of 1,2-hexadecanediol, 12 mmol of oleic acid, 12 mmol of oleylamine, and 40 ml of benzyl ether. That is to say in proportions five times higher for the hexadecanediol compared to the Fe(III) acetylacetonate and six times higher for the surfactants (oleic acid and oleylamine) compared to the Fe(III) acetylacetonate. Thermal controlling is carried out using a thermocouple probe to control the temperature and the duration of the high temperature treatment. The reaction mixture was magnetically stirred and degassed at room temperature for 60 min, then was heated and kept at 100 °C for 30 min to remove water. Subsequently, the temperature was increased and kept at 200 °C, for 30 min, then, heated (to reflux) and kept at 300 °C for 60 min. The final mixture is cooled to room temperature and purified three times with ethanol and hexane. A black magnetic precipitate is obtained after magnetic settling. The precipitate is redispersed in 20 ml of hexane and a ferrofluid composed of surfaced CoFe$_2$O$_4$ nanoparticles is obtained. Figure 1 illustrates the thermal decomposition process. It is interesting to note herein that the presence of the used surfactants helps the good dispersion of the obtained NPs in hexane. However, the presence of hexadecanediol helps to initiate the reaction by promoting the decomposition of the metal precursor's acetylacetonates. The choice of benzyl ether as an appropriate solvent for this process because its boiling temperature (298 °C) is higher than the decomposition temperature of precursors. The equation of the reaction is as follows [23]:

$$Co(C5H7O2)2 + Fe(C5H7O2)3 \rightarrow CoFe2O4 + CH3COCH3 + CO2 \qquad (Eq.2)$$

According to the equation (Eq. 2), in the presence of oleylamine, oleic acid and 1,2-hexadecanediol, thermal decomposition of acetylacetonates of cobalt and iron produced cobalt ferrite nanoparticles, releasing acetone and carbon dioxide as by-products.

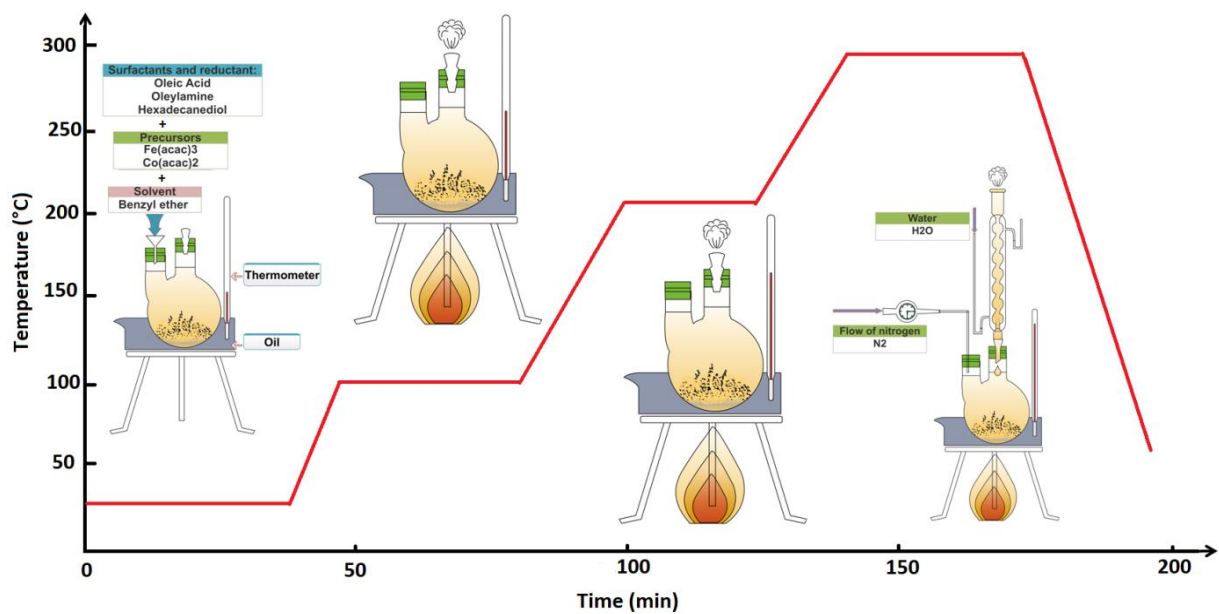

Figure 1: Illustration of thermal decomposition method.

Various synthesis parameters described above in the initial protocol have been modified in order to know the influence of synthesis parameters on the shape and size of nanoparticles: the decomposition temperature, the duration of the heat treatment or the quantity of the reagents. This also allowed to better understand the role of reagents such as hexadecanediol, oleic acid or oleylamine in the synthesis.

*2.3. Characterization techniques*

In order to get information about the mass loss of $CoFe_2O_4$ NPs, thermogravimetric analysis (TGA, TA Instrument Q500) was used to know the percentage and the degradation temperature of the organic molecules on the surface of nanoparticles. The sample was analyzed under an inert atmosphere, the heating rate is 10°C/min, the temperature range is between 25-600°C and the mass used is between 10 and 30 mg.
Fourier Transform - Infrared Spectroscopy (FT-IR) spectra were recorded in the region from 250 to 4000 cm$^{-1}$ by using ABB Bomem FTLA2000 on KBr-dispersed sample pellets. In order to avoid the signal saturation effects, the studied powders are diluted with KBr (transparent to infra-red radiation), and compressed into a disk with a diameter of 1 cm, in the form of pellets consisting of 30 mg of KBr and 1 mg of the sample. The spectra was recorded between 400 and 4000 cm$^{-1}$ and processed using the Win-IR software.

X-ray powder diffraction (XRD) patterns of the nanoparticle assemblies were collected on a Bruker D8 Discover diffractometer under CuK$\alpha_1$ radiation ($\lambda$=1.5406 Å) at 25°C. Scanning angle 2θ ranging from 10° to 100° with a step of 0.1°. The objective of this analysis is the determination of the phases present in the samples, verification of the absence of secondary phases, the calculation of the unit cell parameter as well as the determination of the particle size.

Scanning transmission electron microscopy (STEM) studies and associated energy dispersive X-ray spectroscopy (EDS) microanalysis were performed using a FEI electronic microscopy operating at 30 KV. The nanoparticles were dispersed on holey carbon grids for STEM observation. EDS chemical analysis were also carried out on several zones to determine locally the quantity of the elements.

The magnetic properties of the CoFe$_2$O$_4$ nanoparticles were studied at various temperatures using a Quantum Design MPMS-XL-7CA SQUID magnetometer with a magnetic field strength up to 6 T. The principle of this measurement is based on the displacement of the sample within a set of superconducting detection coil. During the movement of the sample through the coils at a given temperature and magnetic field, the magnetic moment of the sample induces an electric current in the sensing coils. Any change of this current in the detection circuit induces a change of magnetic flux; therefore, by moving the sample on either side of the detection coils, the magnetic flux is integrated. A flux transformer is used to transmit the signal to the SQUID.

## 3.    Results and discussion

Figure 2 represents the mass loss of the synthesized CoFe$_2$O$_4$ NPs as a function of temperature under an inert atmosphere. As can be seen in Thermogravimetric analysis (black curve) and differential thermal analysis (blue curve), a mass loss of about 5% is detected below 300 °C (573 K), which can be attributed to solvent remainders and adsorbed humidity. However, near to 350 °C (623 K) a mass loss of about 12.5% is clearly identified as the thermal degradation of the surfactants (oleic acid and oleylamine) on the surface of the nanoparticles (the boiling point between 250°C (523 K) and 360 °C (633 K)). Moreover, no other peaks are observed in the range of test temperature, which means that there is no phase change of the material after heating at high temperature (T ≤ 600 °C (873 K)). This diagram confirms thermal stability and negligible structure leaching.

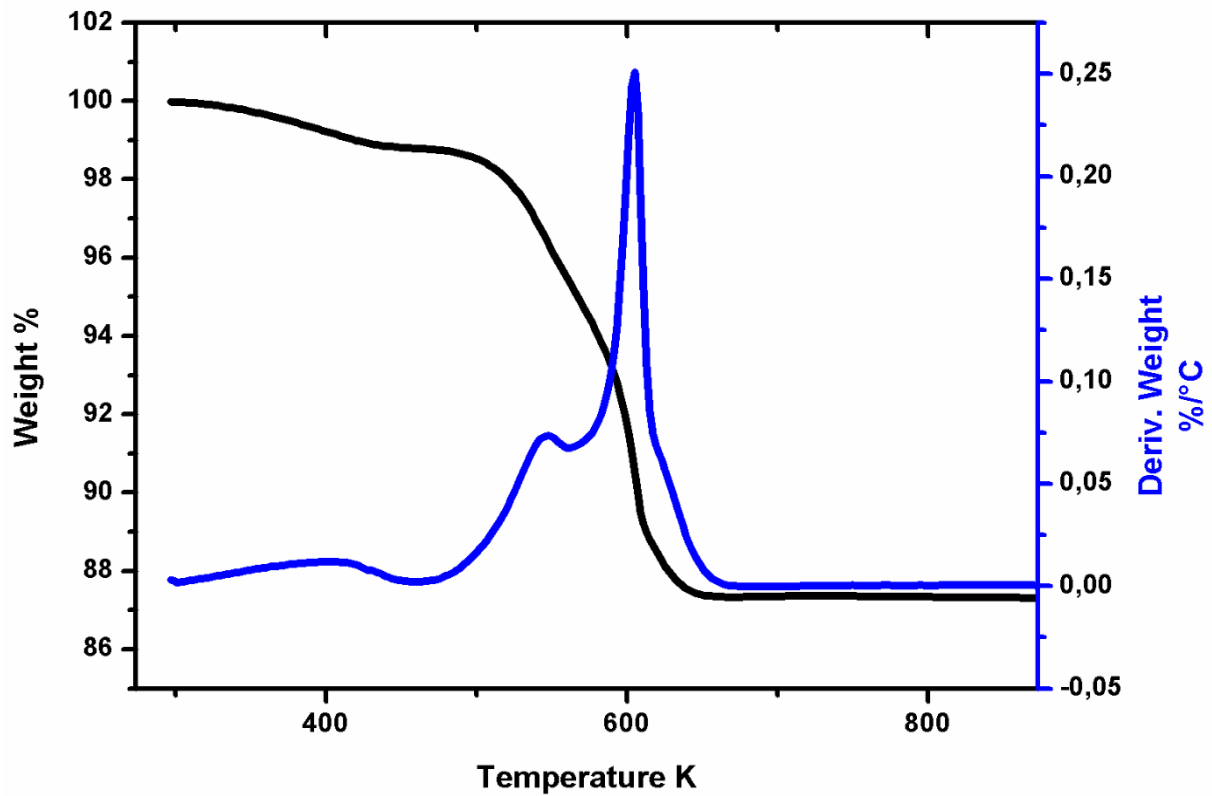

Figure 2: Thermogravimetric analysis (black curve) and differential thermal analysis (blue curve) of the synthesized $CoFe_2O_4$ nanoparticles at atmospheric pressure.

Figure 3 represents the FT-IR analysis to identify the presence of functional groups of organic molecules surrounding the NPs as well as the vibrational modes of metal-oxygen bonds for spinel structure. As shown in the figure, the principal vibrational modes of metal-oxygen (M-O) bonds are present between 300 and 670 cm$^{-1}$ that correspond to metals in a tetrahedral or octahedral configuration for spinel structures. In general, bands of M-O bonds in the octahedral sites appear at 380-450 cm$^{-1}$, whereas they are around 540-600 cm$^{-1}$ for tetrahedral sites. In our case, a band of M-O bonds in the octahedral sites appear at 400 cm$^{-1}$, whereas the band of M-O bonds in the tetrahedral sites appears at 591 cm$^{-1}$.

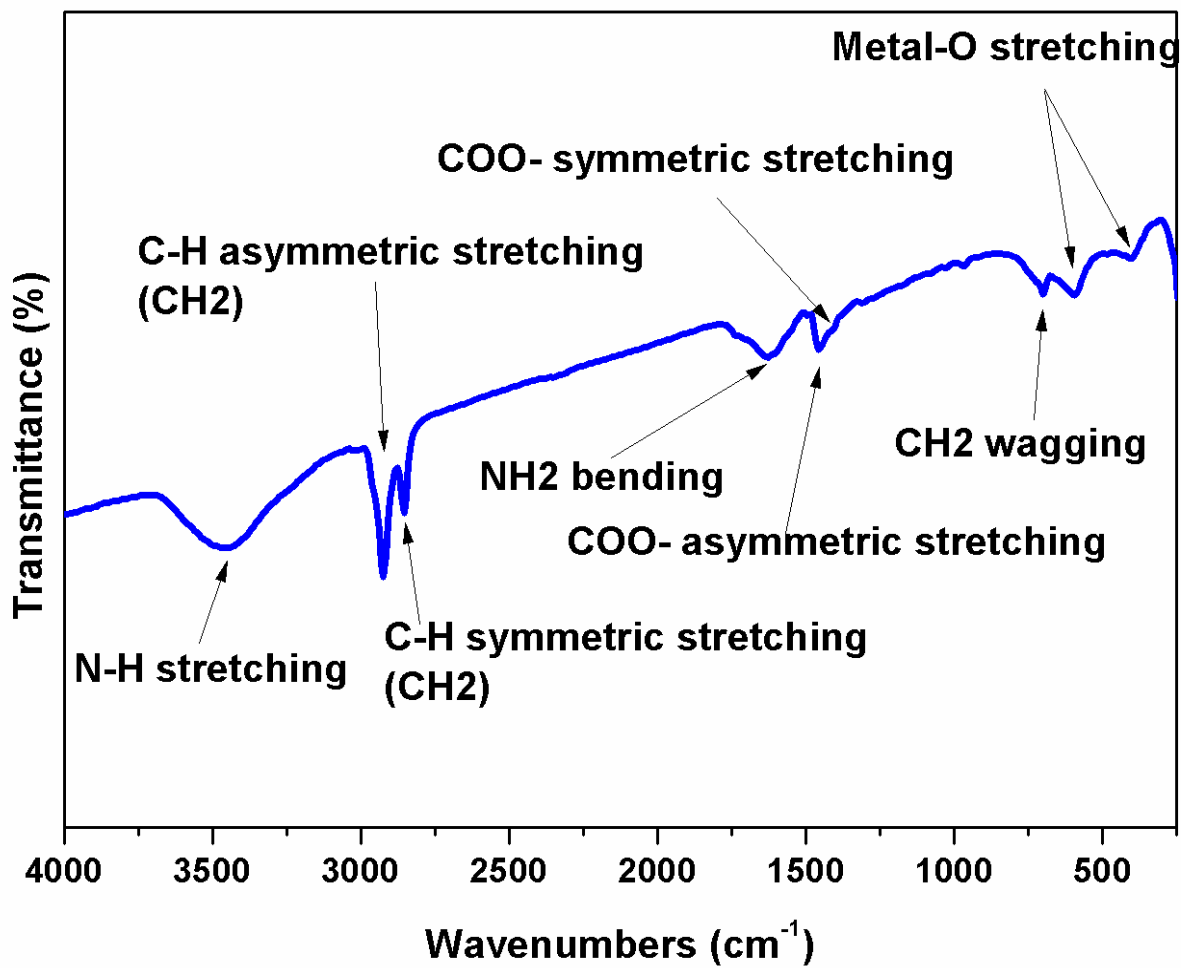

Figure 3: The infrared spectrum of the synthesized $CoFe_2O_4$ nanoparticles.

Further, the absorption bands observed in the range 670-3700 cm$^{-1}$ (see Figure 3) correspond to the vibration bands of the surfactant groups. The bands at 2924 and 2851 cm$^{-1}$ can be assigned to the asymmetrical and symmetrical stretching of $CH_2$ groups, characteristic of the hydrocarbon chain of the used surfactants. Two bands at 1453 and 1408 cm$^{-1}$ are observed and correspond to the asymmetric and symmetric elongations of the carboxylate groups (COO$^-$ stretching). In addition, vibrational modes observed at 3440 and 1611 cm$^{-1}$ correspond to the angular deformations of the amine groups (NH stretching and $NH_2$ bending, respectively), another band appearing at 702 cm$^{-1}$ correspond to $CH_2$ wagging.

In order to verify the spinel structure and to estimate the particle size, X-ray diffraction measurement was carried out. The X-ray diffraction pattern of the synthesized $CoFe_2O_4$ (Figure 4) shows that the obtained diffraction peaks correspond well to the spinel structure (JCPDS No. 04-016-3954) with face-centered cubic phase. Traces of any other phases, kind of detectable impurities or intermediate phase were not observed. It is worthy to note that all other nanoparticles synthesized with a modification of the reaction conditions (see below) led to the similar phase purity and no clear difference could be spotted from the diffractograms

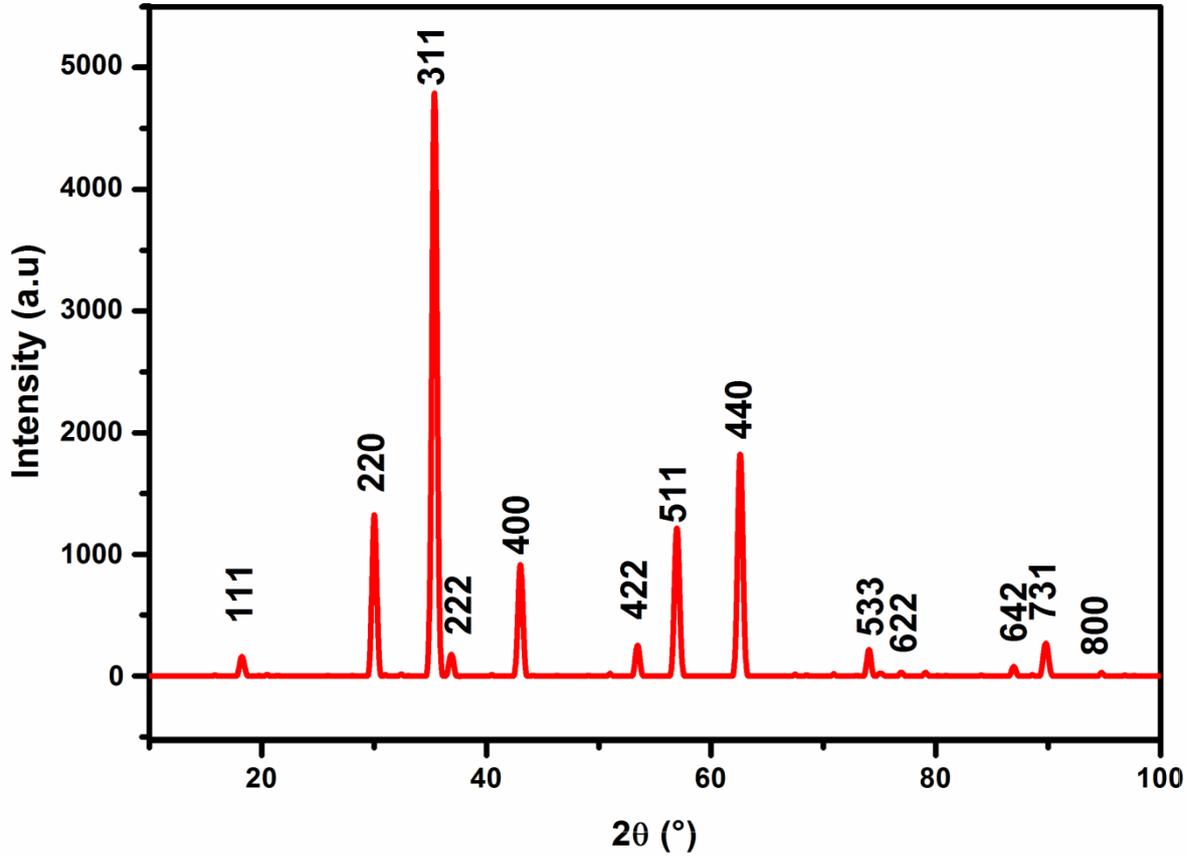

Figure 4: X-ray diffraction pattern of the synthesized CoFe$_2$O$_4$ nanoparticles.

The broad diffraction peaks obtained are expected for such small crystalline domains. The Scherrer's formula allows estimating the crystallite size by taking the full width at half maximum (FWHM) of the main diffraction peak (311).

$$D = \frac{0.9\lambda}{\beta cos(\theta)} \qquad (Eq.3)$$

Where $\lambda$ is the X-ray wavelength, $\beta$ is the broadening at half the maximum intensity (FWHM) of the hkl peak (in our case the 311 peak) and $\theta$ is the Bragg angle of this peak. Generally, β is corrected according to the formula $\sqrt{\beta_x^2 - \beta_{Si}^2}$, where $\beta_x$ is the experimental FWHM and $\beta_{Si}$ is the FWHM of a standard silicon sample. The average crystallite size estimated is 11.2 nm. This value can be used to calculate the specific surface area using the formula : $SSA = \frac{A}{V \cdot \rho}$, where A is the surface area, V is the volume of nanoparticle (sphere in our case) and ρ is the theoretical density of CoFe$_2$O$_4$ which is 5259*10$^3$ g.m$^{-3}$ [24]. The specific surface area of the synthesized CoFe$_2$O$_4$ nanoparticles is 102.04 m$^2$.g$^{-1}$.

According to the images obtained by STEM (Figure 5), the synthesized CoFe$_2$O$_4$ nanoparticles are spherical and highly monodispersed. The statistical analysis by ImageJ (macro particle size analyzer), allowed to obtain the histogram giving rise the information about the size distribution of the nanoparticles. We can deduct from the histogram that the average diameter of NPs is around 10.7 nm.

It is worth noting that the mean size calculated from STEM images is in good agreement with that estimated from XRD. This agreement supports that these nanoparticles are single crystals. The slight difference observed between the two techniques could be explained by the fact that for XRD only the largest particles are counted, whereas in STEM, the size distribution with an average diameter is obtained on a limited number of particles (around 550 NPs).

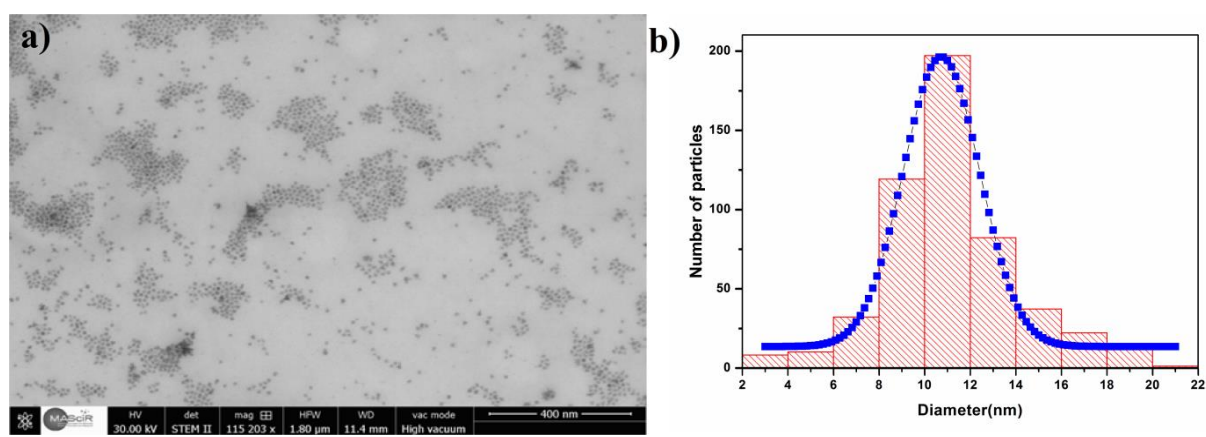

Figure 5: STEM image (a) and size distribution histogram (b) of the synthesized CoFe$_2$O$_4$ nanoparticles.

To verify the formation of CoFe$_2$O$_4$ phase, energy dispersive X-ray spectroscopy (EDS) was used to analyze the chemical composition as shown in table 1. In order to carry out the analysis under good condition, we removed the organic molecules surrounded the surface of the nanoparticles by a heat treatment at 400 °C for 2 hours. EDS results indicated that the ratio of Co/Fe is 1/2, which agree well with the ratio of initial metal precursors. Thus, the final Co/Fe composition could be readily controlled. This conclusion is in good agreement with the result obtained by Lu et al. [23].

| Element | Line s. | Mass [%] | Mass Norm[%] | Atom [%] |
|---------|---------|----------|--------------|----------|
| Oxygen  | K-Serie | 27.52    | 27.83        | 57.84    |
| Iron    | K-Serie | 45.97    | 46.50        | 27.68    |
| Cobalt  | K-Serie | 25.37    | 25.67        | 14.48    |
|         |         | 98.86    | 100.00       | 100.00   |

Table 1: EDS spectra for $CoFe_2O_4$ nanoparticles.

One of the main characteristics of $CoFe_2O_4$ nanoparticles is that they are magnetic. In order to study their magnetic properties, measurements were carried out on $CoFe_2O_4$ NPs with and without organic molecules. Notice that in order to study $CoFe_2O_4$ NPs without organic molecules, we removed the organic molecules surrounded the surface of the nanoparticles by a heat treatment at 400 °C for 2 hours.

Figure 6 shows the magnetization curves as a function of the magnetic field at 300K and 10K of $CoFe_2O_4$ NPs. As it is clearly seen from the plots, a difference in saturation magnetizations of about 12% is spotted between $CoFe_2O_4$ NPs with and without organic molecules. Therefore, the contribution of the organic molecules is estimated at 12%. This value agrees well with that one found by TGA measurement (12.5%).

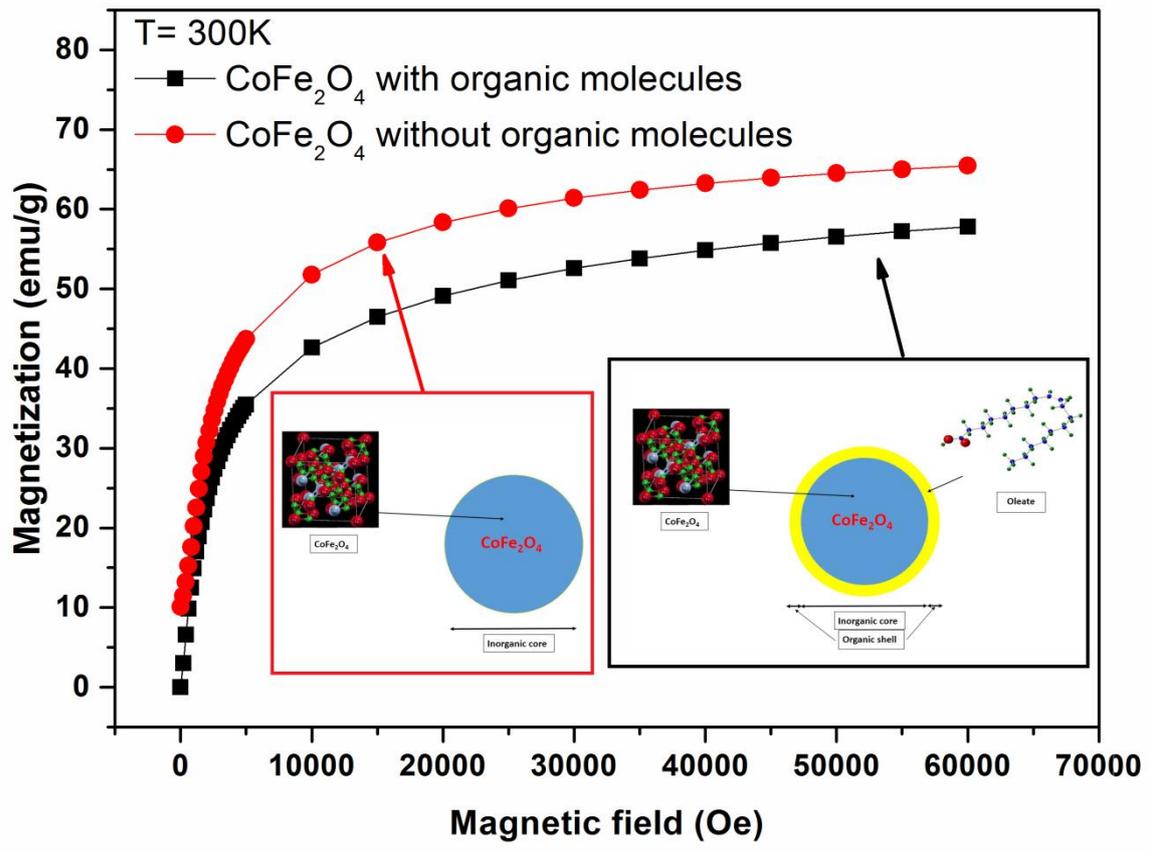
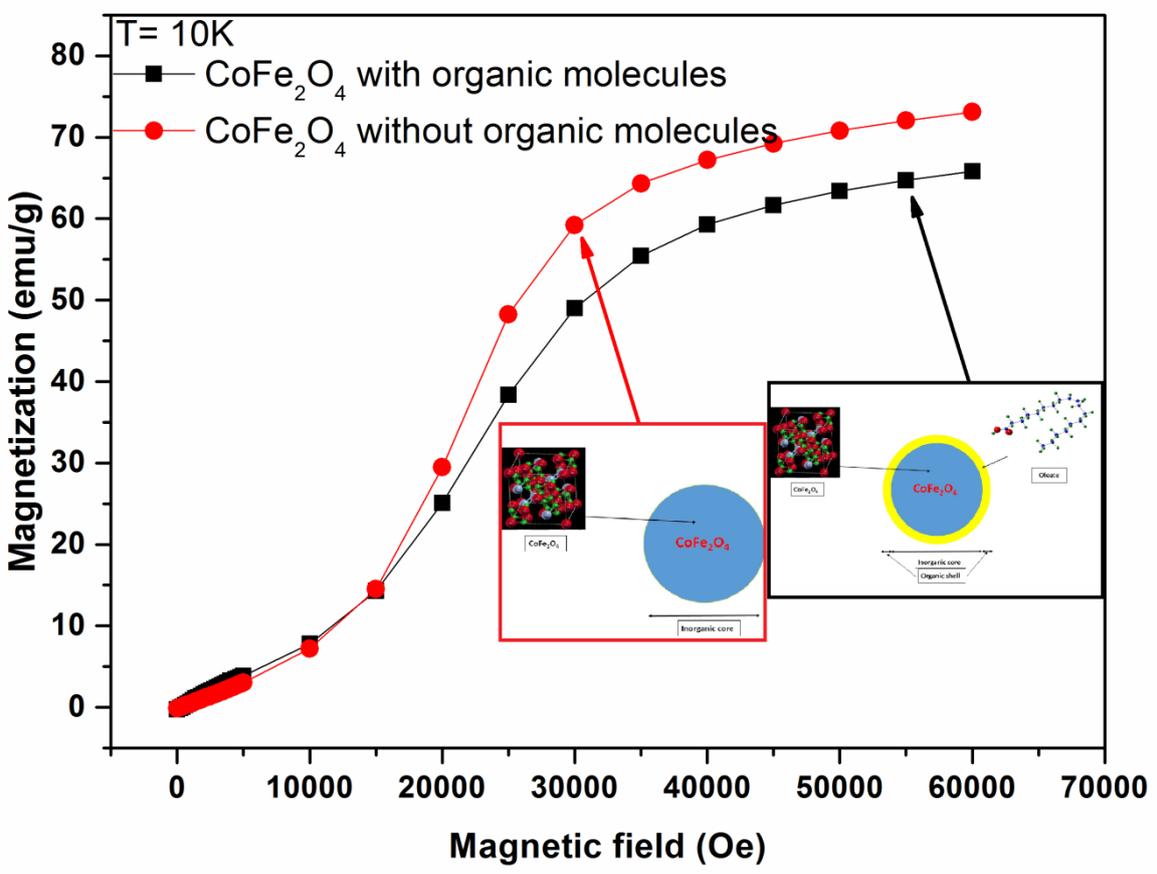

Figure 6: Magnetization as a function of magnetic field of $CoFe_2O_4$ nanoparticles with and without magnetic field at 10 K (bottom) and 300 K (top).

The transition from the superparamagnetic state to the blocked state takes place at a temperature called blocking temperature (TB). This depends on the material, the size of the particles and also on the presence of interparticle magnetic interactions. To determine ($T_B$), the evolution of the magnetization as a function of the temperature is performed under a constant magnetic field of 100 Oe. Figure 7 shows the zero-field-cooled (ZFC) and field-cooled (FC) curves of cobalt ferrite nanoparticles. The $CoFe_2O_4$ NPs have a blocking temperature at the vicinity of 300K, which is close to the room temperature. The difference between ZFC and FC magnetizations below $T_B$ is caused by the energy barriers of the magnetic anisotropy [25]. The magnetic anisotropy constant K of the $CoFe_2O_4$ nanoparticles can be estimated using the formula $K = 25 K_B T_B V^{-1}$, where $K_B$ is the Boltzmann constant, $T_B$ is the blocking temperature of the samples, and V is the volume of a single particle. The calculated magnetic anisotropy constant K of our sample is found equal to $1.6*10^5$ J.m$^{-3}$. This estimated value is comparable with the value reported in the literature for the bulk [8].

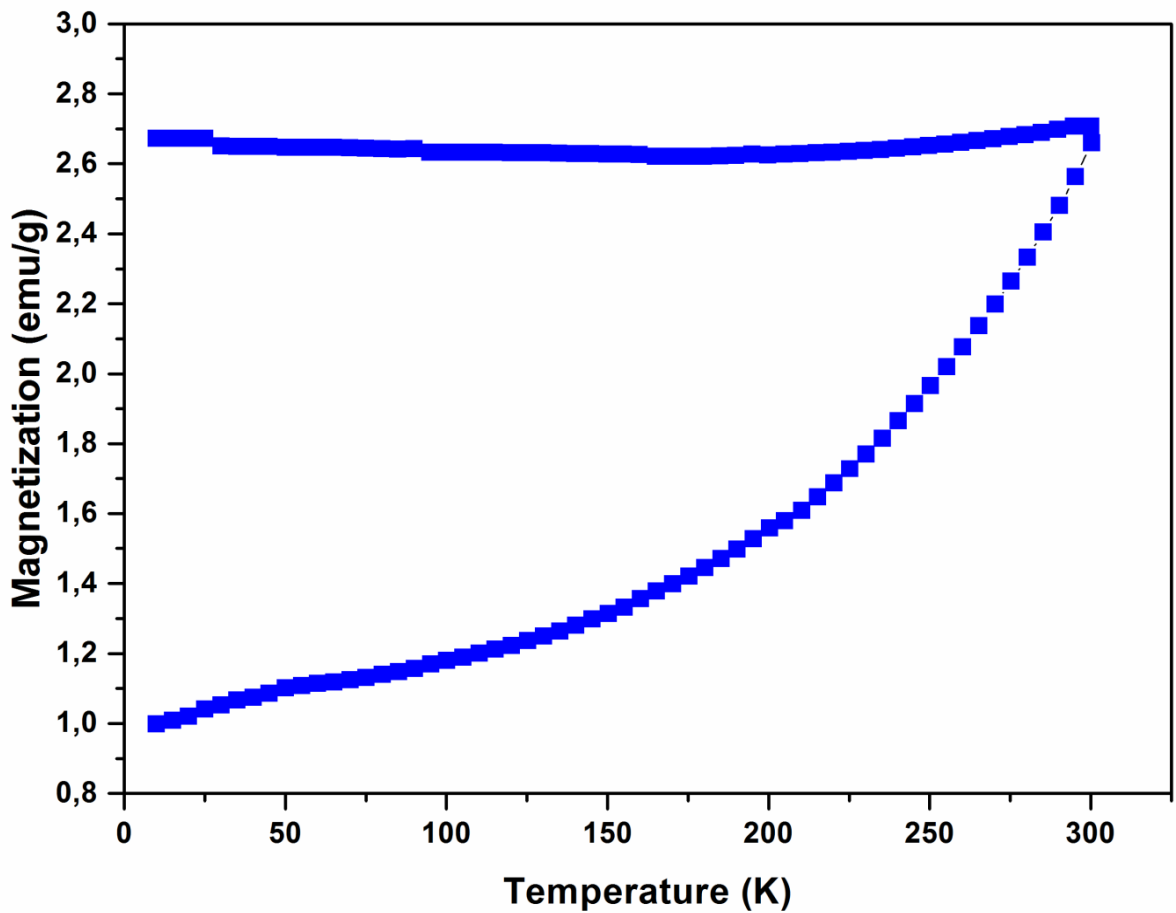

Figure 7: ZFC/FC curves of cobalt ferrite nanoparticles measured at temperatures ranging from 10 K to 300 K and with an applied magnetic field of 100 Oe.

At room temperature (300 K), the behavior of dispersed nanoparticles is superparamagnetic (see Figure 8), their magnetization curve is reversible. At low temperature, around 10 K, the ferrofluid contains nanoparticles is frozen and the intrinsic characteristics of the nanoparticles are found. The magnetization curve exhibits a large hysteresis loop, similar to a hard magnetic material, and with a coercivity of about 18.6 kOe, suggesting the presence of particles in a blocked and non-equilibrium state. In contrast, the coercivity value of $CoFe_2O_4$ NPs is only 315 Oe at 300 K due to the additional thermal activation energy which decreasing the exchange interaction between spin moment. At 10 K, the coercivity values are in the same order of magnitude as those of $CoFe_2O_4$ nanotubes [14] and nanowires [26] fabricated by electrospinning, nanorods synthesized by microemulsion [27] and nanoparticles synthesized by co-precipitation method [28].

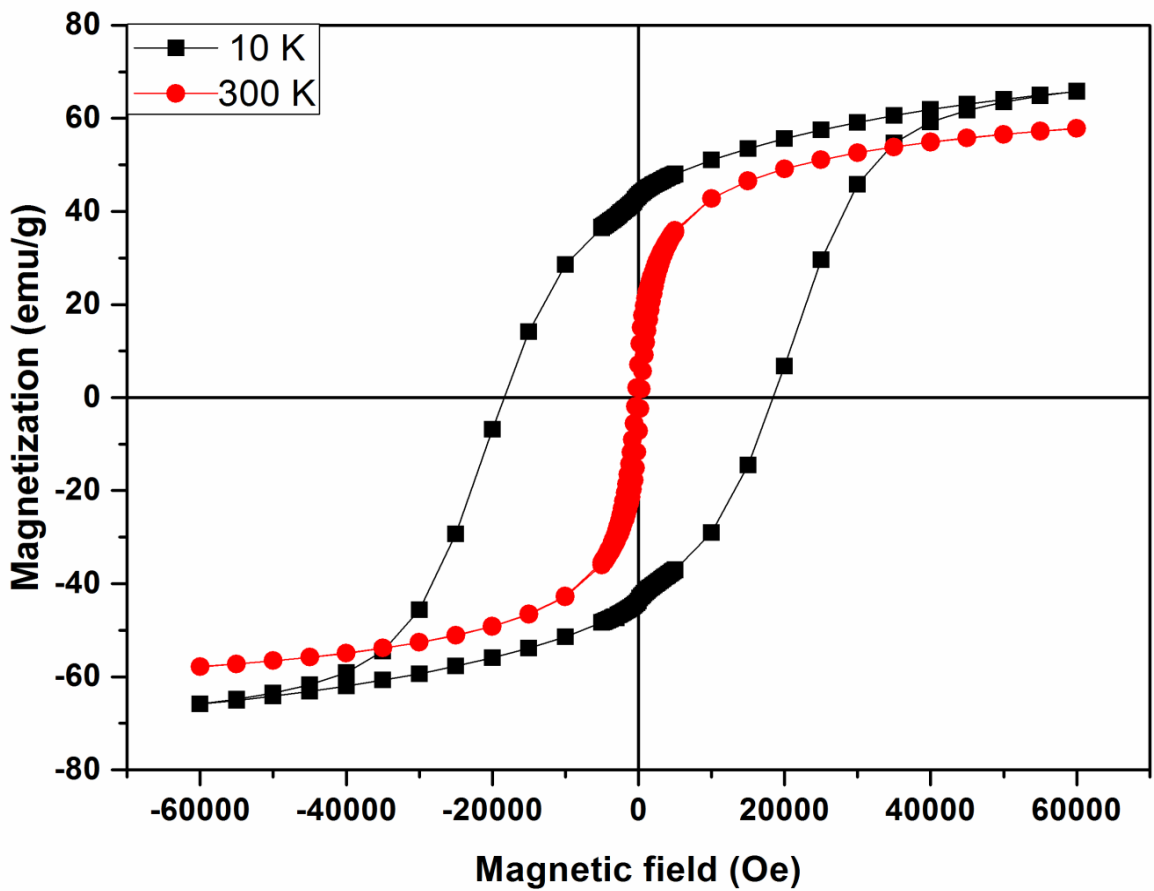

Figure 8: Hysteresis loops of the $CoFe_2O_4$ nanoparticles measured at 10 K and 300 K.

Finally, the values of saturation magnetization, *Ms*, obtained for $CoFe_2O_4$ NPs in this work, are in the range of 58 to 65 emu/g. These values are slightly smaller than those of the bulk $CoFe_2O_4$ ranging between 80 and 85 emu/g [8, 13]. This magnetization reduction may be explained by the magnetic moment disorder at the particle surface.

It is interesting to note herein that the size and the shape of the obtained particles depend on various parameters such as the quantity of precursors, the temperature or the duration of the heat treatment [13, 18]. More than that, they depend on the oleic acid/ oleylamine ratio and the presence or absence of hexadecanediol.

The fact of the matter, when the amount of the surfactants increases four times compared to the initial protocol, the obtained nanoparticles are smaller (*d* = 7.3 ± 2.1 nm), but they are polydisperse as seen in figure 9.(a). Another parameter likely to reduce the particle size is to work under more dilute conditions [29]. For example, a twofold dilution compared to the original protocol, leading to a diameter comparable to the first one (d = 7.8 ± 1.6 nm), but with a lower polydispersity of NPs than those obtained by increasing the amount of surfactants (see Figure 9.b). On the contrary, for increasing the particle size, we have increased the duration of the heat treatment, Figure 9.c shows that the size of $CoFe_2O_4$ NPs increased from 10.7 nm to 13.2 nm when the duration of heat treatment increased from 60 min to 120 min. This result is in good agreement with the work of Perez-Mirabet et al. [30]. In the meanwhile, another parameter could involve in the control of NPs size, is the nature of the solvent used during the reaction. Baaziz et al. [31] carried out the synthesis with both polar and non-polar solvents having different boiling temperatures. With the non-polar solvents, the authors found that the size of the nanoparticles increased almost linearly when the boiling temperature of the solvent increased, they suggested that the growth step of the particles depends on the temperature of the reaction. However, using the polar solvents, the size of the nanoparticles deviated from this linear growth. The authors concluded that the nature of the solvent has an influence on the nucleation and growth steps of the nanoparticles. That was related to the stability of the formed metal complexes which depends on the interactions with the solvent and its functional group. Basing in these conclusions, we replaced the benzyl ether (*Bp*= 298 °C) by octadecene (*Bp*= 318 °C) to see the influence of the solvent on size and morphology of NPs. We found that the nanoparticles obtained in octadecene have a higher size than those obtained before in benzyl ether. This result confirms the conclusion of Baaziz et al. [31] which says that the nature of the solvent has an influence on the nucleation and growth steps of nanoparticles.

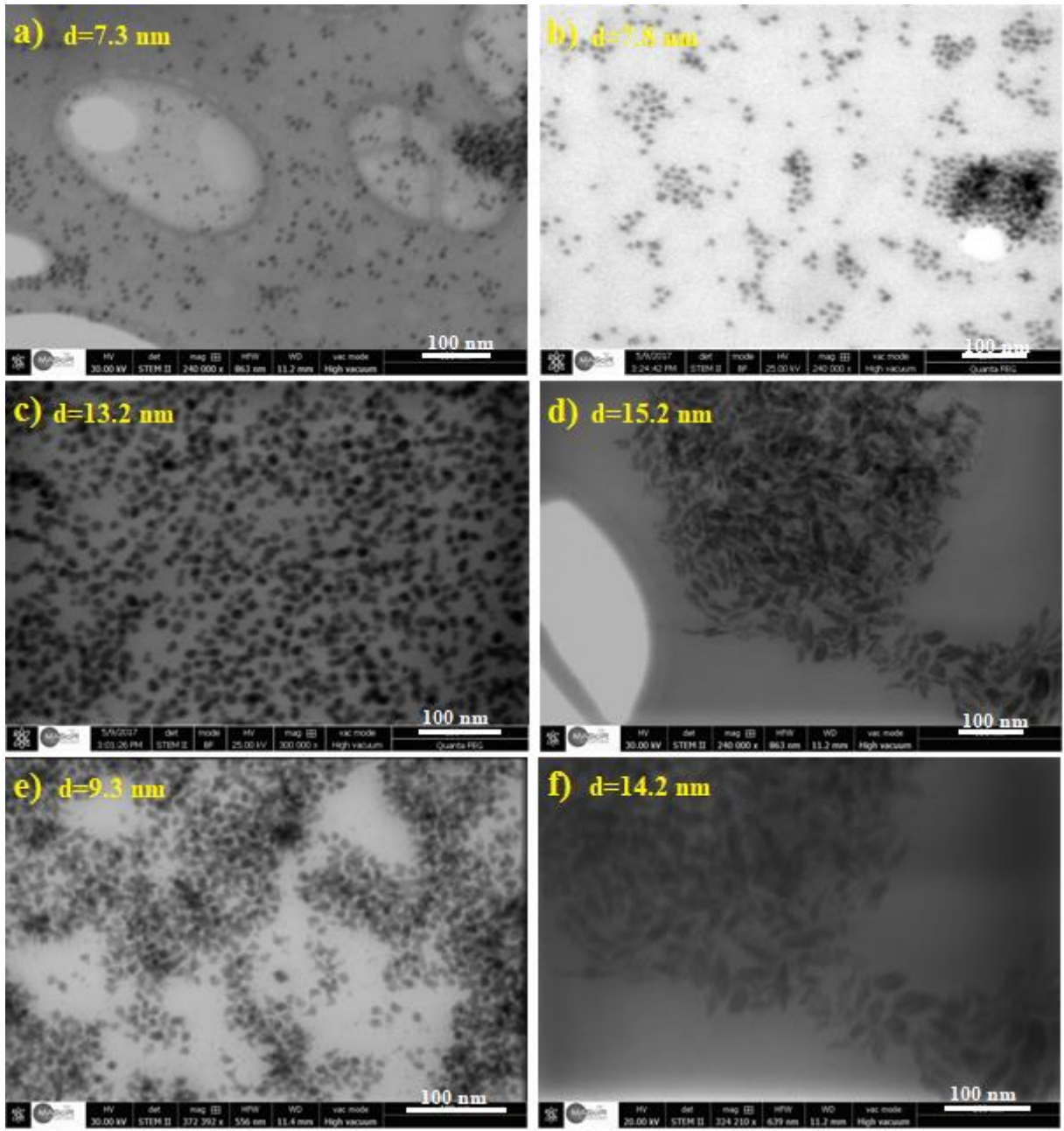

Figure 9: STEM image of the synthesized CoFe$_2$O$_4$ nanoparticles.

Concerning the indispensability of the hexadecanediol, Crouse et al. [32] reported that the absence of hexadecanediol does not influence the particle size, but affects NPs polydispersity. The higher is the concentration of hexadecanediol, the greater the particle size distribution is, and with a linear dependence. In the other hand, Moya et al. [33] reported that hexadecanediol favors the decomposition of acetylacetonate precursors, and therefore nucleation of the particles at lower temperatures. So, the CoFe$_2$O$_4$ nanoparticles obtained without hexadecanediol have a defective crystallographic structure. In this context, we have synthesized CoFe$_2$O$_4$ NPs without hexadecanediol, the STEM image (Figure 9.d) shows that the synthesized particles without hexadecanediol are slightly larger, more polydisperse than those synthesized with hexadecanediol, and have a poorly defined morphology; which is different from the results described above. The role of hexadecanediol in the synthesis is not clearly defined; complementary supporting evidence are needed to shed more light on its influence. Continuously, in order to study the role of each surfactant in the synthesis, $CoFe_2O_4$ NPs were synthesized in the absence of each one. Without oleic acid, the particles are very small and more aggregated (Figure 9.e), whereas without oleylamine the particles have a poorly defined morphology and are highly polydisperse (Figure 9.f). We surmise then that oleic acid is a surfactant that stabilizes nanoparticles, and oleylamine provides the basic medium necessary to form oxides of the spinel structure.

## 4.    Conclusions

In summary, the structural and magnetic properties of $CoFe_2O_4$ nanoparticles are presented. The thermal decomposition process allowed us to obtain spherical and monodisperse cobalt ferrite NPs surfaced by organic molecules and stabilized in an organic solvent. Using STEM analysis, we found that their size and shape could be controlled by varying certain parameters such as the synthesis temperature, the quantity, and nature of reagents.  EDS and XRD measurements confirmed the formation of $CoFe_2O_4$ nanoparticles with spinel structure.

The magnetic investigations revealed a blocking temperature very close to the room temperature, attesting then the room temperature superparamagnetic behavior of the $CoFe_2O_4$ NPs with a small coercivity value of about 315 Oe. Otherwise, At 10 K, $CoFe_2O_4$ nanoparticles show the intrinsic characteristic behavior similarly to $CoFe_2O_4$ nanotubes, nanowires or nanorods.

The results obtained in this work are likely to offer useful information about the preparation and the role of different parameters in this synthesis route of $CoFe_2O_4$ NPs, promising for application in magnetic nanodevices and biomagnetic applications.

**Acknowledgments**
This work has been carried out with the support of the Ministry of Higher Education, Scientific Research,and Professional Training (Enssup) (Morocco) and the National Center for Scientific and Technological Research(CNRST) through the grant Number: PPR15, and by the European H2020-MC-RISE-ENIGMA action (N°778072) and FEDER.

# References

1. Frey NA, Peng S, Cheng K, Sun S (2009) Magnetic nanoparticles: synthesis, functionalization, and applications in bioimaging and magnetic energy storage. Chem Soc Rev 38:2532. https://doi.org/10.1039/b815548h

2. Lu A-H, Salabas EL, Schüth F (2007) Magnetic Nanoparticles: Synthesis, Protection, Functionalization, and Application. Angew Chem Int Ed 46:1222–1244. https://doi.org/10.1002/anie.200602866

3. Raj K, Moskowitz R (1990) Commercial applications of ferrofluids. J Magn Magn Mater 85:233–245. https://doi.org/10.1016/0304-8853(90)90058-X

4. Anwar H, Maqsood A (2014) Comparison of structural and electrical properties of $Co^{2+}$ doped Mn-Zn soft nano ferrites prepared via coprecipitation and hydrothermal methods. Mater Res Bull 49:426–433. https://doi.org/10.1016/j.materresbull.2013.09.009

5. Huang X, Li Y, Li Y, et al (2012) Synthesis of PtPd Bimetal Nanocrystals with Controllable Shape, Composition, and Their Tunable Catalytic Properties. Nano Lett 12:4265–4270. https://doi.org/10.1021/nl301931m

6. Rebrov EV, Gao P, Verhoeven TMWGM, et al (2011) Structural and magnetic properties of sol–gel $Co_{2x}Ni_{0.5-x}Zn_{0.5-x}Fe_2O_4$ thin films. J Magn Magn Mater 323:723–729. https://doi.org/10.1016/j.jmmm.2010.10.031

7. Lu Z, Gao P, Ma R, et al (2016) Structural, magnetic and thermal properties of one-dimensional $CoFe_2O_4$ microtubes. J Alloys Compd 665:428–434. https://doi.org/10.1016/j.jallcom.2015.12.262

8. Nam PH, Lu LT, Linh PH, et al (2018) Polymer-coated cobalt ferrite nanoparticles: synthesis, characterization, and toxicity for hyperthermia applications. New J Chem 42:14530–14541. https://doi.org/10.1039/C8NJ01701H

9. Kumar L, Kumar P, Kar M (2013) Cation distribution by Rietveld technique and magnetocrystalline anisotropy of Zn substituted nanocrystalline cobalt ferrite. J Alloys Compd 551:72–81. https://doi.org/10.1016/j.jallcom.2012.10.009

10. Yao L, Xi Y, Xi G, Feng Y (2016) Synthesis of cobalt ferrite with enhanced magnetostriction properties by the sol−gel−hydrothermal route using spent Li-ion battery. J Alloys Compd 680:73–79. https://doi.org/10.1016/j.jallcom.2016.04.092

11. Wang YC, Ding J, Yin JH, et al (2005) Effects of heat treatment and magnetoannealing on nanocrystalline Co-ferrite powders. J Appl Phys 98:124306. https://doi.org/10.1063/1.2148632

12. Yan L, Wang Y, Li J, et al (2008) Nanogrowth twins and abnormal magnetic behavior in $CoFe_2O_4$ epitaxial thin films. J Appl Phys 104:123910. https://doi.org/10.1063/1.3033371

13. Song Q, Zhang ZJ (2004) Shape Control and Associated Magnetic Properties of Spinel Cobalt Ferrite Nanocrystals. J Am Chem Soc 126:6164–6168. https://doi.org/10.1021/ja049931r

14. Fu J, Zhang J, Peng Y, et al (2012) Unique magnetic properties and magnetization reversal process of $CoFe_2O_4$ nanotubes fabricated by electrospinning. Nanoscale 4:3932. https://doi.org/10.1039/c2nr30487b

15. Mahhouti Z, Ben Ali M, El Moussaoui H, et al (2016) Structural and magnetic properties of $Co_{0.7}Ni_{0.3}Fe_2O_4$ nanoparticles synthesized by sol–gel method. Appl Phys A 122:. https://doi.org/10.1007/s00339-016-0178-5

16. Galarreta I, Insausti M, Gil de Muro I, et al (2018) Exploring Reaction Conditions to Improve the Magnetic Response of Cobalt-Doped Ferrite Nanoparticles. Nanomaterials 8:63. https://doi.org/10.3390/nano8020063

17. Asgarian SM, Pourmasoud S, Kargar Z, et al (2018) Investigation of positron annihilation lifetime and magnetic properties of $Co_{1-x}Cu_xFe_2O_4$ nanoparticles. Mater Res Express 6:015023. https://doi.org/10.1088/2053-1591/aae55d

18. Sobhani-Nasab A, Behpour M, Rahimi-Nasrabadi M, et al (2019) New method for synthesis of $BaFe_{12}O_{19}/Sm_2Ti_2O_7$ and $BaFe_{12}O_{19}/Sm_2Ti_2O_7/Ag$ nano-hybrid and investigation of optical and photocatalytic properties. J Mater Sci Mater Electron 30:5854–5865. https://doi.org/10.1007/s10854-019-00883-3

19. Yang C, Yan H (2012) A green and facile approach for synthesis of magnetite nanoparticles with tunable sizes and morphologies. Mater Lett 73:129–132. https://doi.org/10.1016/j.matlet.2012.01.031

20. Li X-H, Xu C-L, Han X-H, et al (2010) Synthesis and Magnetic Properties of Nearly Monodisperse $CoFe_2O_4$ Nanoparticles Through a Simple Hydrothermal Condition. Nanoscale Res Lett 5:1039–1044. https://doi.org/10.1007/s11671-010-9599-9

21. Aparna ML, Grace AN, Sathyanarayanan P, Sahu NK (2018) A comparative study on the supercapacitive behaviour of solvothermally prepared metal ferrite ($MFe_2O_4$, M = Fe, Co, Ni, Mn, Cu, Zn) nanoassemblies. J Alloys Compd 745:385–395. https://doi.org/10.1016/j.jallcom.2018.02.127

22. Hyeon T, Lee SS, Park J, et al (2001) Synthesis of Highly Crystalline and Monodisperse Maghemite Nanocrystallites without a Size-Selection Process. J Am Chem Soc 123:12798–12801. https://doi.org/10.1021/ja016812s

23. Lu LT, Dung NT, Tung LD, et al (2015) Synthesis of magnetic cobalt ferrite nanoparticles with controlled morphology, monodispersity and composition: the influence of solvent, surfactant, reductant and synthetic conditions. Nanoscale 7:19596–19610. https://doi.org/10.1039/C5NR04266F

24. Nlebedim IC, Jiles DC (2014) Dependence of the magnetostrictive properties of cobalt ferrite on the initial powder particle size distribution. J Appl Phys 115:17A928. https://doi.org/10.1063/1.4867343

25. Gräf CP, Birringer R, Michels A (2006) Synthesis and magnetic properties of cobalt nanocubes. Phys Rev B 73:. https://doi.org/10.1103/PhysRevB.73.212401

26. Wang Z, Liu X, Lv M, et al (2008) Preparation of One-Dimensional $CoFe_2O_4$ Nanostructures and Their Magnetic Properties. J Phys Chem C 112:15171–15175. https://doi.org/10.1021/jp802614v

27. Zhang Z, Rondinone AJ, Ma JX, et al (2005) Morphologically Templated Growth of Aligned Spinel $CoFe_2O_4$ Nanorods. Adv Mater 17:1415–1419. https://doi.org/10.1002/adma.200500009

28. El Moussaoui H, Mahfoud T, Ben Ali M, et al (2016) Experimental studies of neodymium ferrites doped with three different transition metals. Mater Lett 171:142–145. https://doi.org/10.1016/j.matlet.2016.02.072

29. Hara S, Aisu J, Kato M, et al (2018) One-pot synthesis of monodisperse $CoFe_2O_4$@Ag core-shell nanoparticles and their characterization. Nanoscale Res Lett 13:. https://doi.org/10.1186/s11671-018-2544-z

30. Pérez-Mirabet L, Solano E, Martínez-Julián F, et al (2013) One-pot synthesis of stable colloidal solutions of $MFe_2O_4$ nanoparticles using oleylamine as solvent and stabilizer. Mater Res Bull 48:966–972. https://doi.org/10.1016/j.materresbull.2012.11.086

31. Baaziz W, Pichon BP, Fleutot S, et al (2014) Magnetic Iron Oxide Nanoparticles: Reproducible Tuning of the Size and Nanosized-Dependent Composition, Defects, and Spin Canting. J Phys Chem C 118:3795–3810. https://doi.org/10.1021/jp411481p

32. Crouse CA, Barron AR (2008) Reagent control over the size, uniformity, and composition of Co–Fe–O nanoparticles. J Mater Chem 18:4146. https://doi.org/10.1039/b806686h

33. Moya C, Morales M del P, Batlle X, Labarta A (2015) Tuning the magnetic properties of Co-ferrite nanoparticles through the 1,2-hexadecanediol concentration in the reaction mixture. Phys Chem Chem Phys 17:13143–13149. https://doi.org/10.1039/C5CP01052G